\documentclass[11pt]{article}

\usepackage{amstex,sint, epsfig}
\begin{document}

\begin{flushright}
   IUHET-353 \\
   December 1996
\end{flushright}
\vskip 0.5cm
\begin{center}
  {\Large\bf  Krylov space solvers for shifted linear systems}
\end{center}
\vskip 0.5 cm
\begin{center}
{Beat Jegerlehner}
\vskip 0.1cm
Indiana University, Department of Physics\\
Bloomington, IN 47405, U.S.A.
\vskip 0.3cm
\end{center}
\begin{abstract}
We investigate the application of Krylov space methods to the solution of
shifted linear systems of the form $(A+\sigma) x - b = 0$ for several values
of $\sigma$ simultaneously, using only as many matrix-vector operations as
the solution of a single system requires. We find a suitable description of the
problem, allowing us to understand known algorithms in a common framework
and developing shifted methods basing on short recurrence methods,
most notably the CG and the BiCGstab solvers. 
The convergence properties of these shifted solvers are well understood
and the derivation of other shifted solvers is easily possible. The application
of these methods to quark propagator calculations in quenched QCD 
using Wilson and Clover fermions is discussed and numerical examples in 
this framework are presented. With the shifted CG method an optimal 
algorithm for staggered fermions is available. 
\end{abstract}

\section{Introduction}

In many cases Krylov space solvers are the methods of choice for the 
inversion of large sparse matrices. While most Krylov space solvers are 
parameter free and do not have to be tuned to a particular problem, 
exploiting special algebraic properties of the matrix can lead to considerable
acceleration of these algorithms. A recently discussed example is given by
$J$-hermitean matrices, e.g. $JM = M^\dagger J$, where the number of 
matrix-vector products of algorithms like QMR or BiCG can be reduced by a 
factor of two \cite{qmr,deforcrandg5} if multiplications by $J$ and $J^{-1}$ 
are cheap.
Another case which has been discussed in some detail recently
is the application of Krylov space solvers to shifted equations, i.e.
where the solution to
\begin{equation}
(A+\sigma) x - b = 0
\end{equation}
has to be calculated for a whole set of values of $\sigma$. This kind of
problem arises in quark propagator calculations for QCD as well as
other parts of computational physics (see \cite{qmr}).
It has been realized that several algorithms allow one to perform this task 
using only
as many matrix-vector operations as the solution of the most 
difficult single system requires. This has been achieved for the 
QMR \cite{qmr}, the MR \cite{m3r} and the Lanczos-implementation
of the BiCG method \cite{borici}. We present here a unifying discussion
of the principles to construct such algorithms and succeed in constructing
shifted versions of the CG, CR, BiCG and BiCGstab algorithms, using only two 
additional vectors for each mass value. The method is also easily applicable to
many other cases. The key to this construction is the observation that
shifted polynomials, defined by
\begin{equation}
P^\sigma_n(A+\sigma) = c P_n(A)
\end{equation}
where $P_n(A)$ is the polynomial constructed in the Krylov space method, 
are often still useful objects. Since vectors generated by these shifted
polynomials are simply scaled vectors of the original vectors, they are
easily accessible. 

In the following sections we discuss the properties and construction of
shifted polynomials in several cases. We then present the shifted versions
of the above mentioned algorithms and finally perform some numerical 
tests.

\section{Shifted polynomials}

Our ultimate goal is to 
construct an algorithm for a whole trajectory of matrices $A+\sigma$ while
only applying the matrix-vector operations for the inversion of one matrix,
presumably the one with the slowest convergence. In the class of Krylov space
solvers, one deals with residuals or iterates 
which are in some ways derived from polynomials $P_n(z)$ of the matrix $A$:
\begin{equation}
v_n = P_n(A) v_0 .
\end{equation}
We simply define the shifted polynomial $P^\sigma_n(z)$ as
\begin{equation} \label{shiftc}
P_n^\sigma(z+\sigma) = c_n^\sigma P_n(z) .
\end{equation}
$c_n^\sigma$ is determined by the normalization conditions for $P_n(z)$ 
required in the algorithm. 
It is easy to see that we can construct solvers which generate iterates
of the form
\begin{equation}
v_n^\sigma = P_n^\sigma(A) v_0
\end{equation}
without additional matrix-vector products for multiple values of $\sigma$
since the calculation of
\begin{equation}
(A+\sigma) v_n^\sigma = c_n^\sigma (Av_n + \sigma v_n)
\end{equation}
can be derived from matrix-vector products of one single system. 
If $P_n(z)$ is a polynomial which reduces the vector $v_0$, e.g. which is an
approximation to $0$ in some complex region containing the relevant eigenvalues
of $A$ and $c_n^\sigma \le 1$, $P^\sigma_n(z)$ will be a useful polynomial,
too. 

Another class of useful polynomials are the Leja-polynomials, where the
roots of the polynomial are given by the Leja points $z_i$ of a compact
set $K$ in the complex plane not containing the origin 
implicitely defined by
\begin{align}
|z_1| &= \sup_{z \in K} |z| \\
\prod_{k=1}^{j-1} |z_j - z_k| &= \sup_{z\in K} \prod_{k=1}^{j-1} |z - z_k|,
\quad z_j \in K, \quad j=2,\cdots,n .
\end{align}
The Leja points are usually not uniquely defined. The polynomial defined by
\begin{equation}
L_n(z) = \prod_{i=1}^n (1 - z_i^{-1} z)
\end{equation}
is a good approximation of $0$ in $K$. The application of Leja polynomials to 
matrix inversion problems has been described in \cite{leja}. 
If $z_1$ is translation 
invariant, e.g.
\begin{equation}
z_1^{K+\sigma} = z_1^K + \sigma,
\end{equation}
which is for example 
true if K is a circle with center on the positive real axis 
and $\sigma$ is real and positive, all Leja points are translation invariant
and the shifted polynomial is exactly the 
Leja polynomial for the translated region $K$. The application of Leja 
polynomials to construct Krylov space methods for the Wilson matrix is 
currently under investigation.

In the case of formally orthogonal polynomials, which are usually
generated in CG and Lanczos-type algorithms, we can also see that 
the shifted polynomials are exactly the polynomials generated by the
process for the shifted matrix. To see this, we introduce
\begin{equation}
{\cal K}_n(A,v_0) = {\rm span}\{ A^i v_0, i=0\cdots n-1 \} .
\end{equation}
The Lanczos polynomials $Z_n(z)$ have the property of formal orthogonality, 
namely
\begin{equation}
Z_n(A) v_0 \perp {\cal K}_n(A, v_0)
\end{equation}
or, for the non-hermitean process,
\begin{equation}
Z_n(A) v_0 \perp {\cal K}_n(A^\dagger, w_0) 
\end{equation}
for some vector $w_0$. 
It should be noted that usually $Z_n(z)$ is uniquely determined up to
a scalar constant (in the case it is not uniquely determined the Lanzcos 
process can break down \cite{qmr3}). Since 
\begin{equation}
{\cal K}_n(A, v_0) = {\cal K}_n(A+\sigma, v_0),
\end{equation}
we must have
\begin{equation}
Z^\sigma_n(z+\sigma) = \zeta_n^\sigma Z_n(z),
\end{equation}
 since $Z^\sigma_n(z+\sigma)$ is a formally orthogonal polynomial for $A$ 
as well. We therefore expect that the polynomials generated in CG and 
Lanczos-type algorithms are of a shifted structure. We can indeed generate 
the exact processes for several values of $\sigma$ using only one
matrix-vector operation each iteration by calculating the shifted polynomials.

In the following we will show how to calculate the parameters of the shifted
polynomials from the original process in the case of the above mentioned
recurrence relations.

\subsection{Two-term recurrences}

This recurrence is found in MR-type methods or in hybrid methods using
MR-type iterations. We assume here that the leading coefficient is one. The
polynomial is given directly as a product of its linear factors:
\begin{equation}
R_n(z) = \prod_{i=1}^n (1-\chi_n z) .
\end{equation}
To calculate the shifted polynomial, we look at a linear factor 
\begin{equation}
(1 - (z+\sigma) \chi') = c (1 - z \chi)
\end{equation}
resulting in
\begin{equation}
\chi' = {\chi\over 1+\sigma\chi}
\end{equation}
and
\begin{equation}
c = {1\over 1+\sigma\chi} .
\end{equation}
The shifted polynomial is therefore given by
\begin{align}
R^\sigma_n(z) &= \prod_{i=1}^n {1\over 1+\sigma\chi_i}
 \left( 1 - {\chi_i\over 1+\sigma \chi_i} z\right) \\
  & =  \rho^\sigma_n R_n(z-\sigma) \\
\rho_n^\sigma &= \prod_{i=1}^n {1\over 1+\sigma\chi_i} .
\end{align}
If the spectrum of the matrix lies in the right half of the complex plane
we can expect that all inverses of the roots lie there, too. 
We can then easily see that
$c_n^\sigma > 1$ for $\sigma > 0$, 
so that the shifted polynomial converges better than the
original polynomial with a rate growing with $\sigma$. 
This is not surprising since we expect the condition 
number of the matrix $A+\sigma$ to decrease for $\sigma>0$. 

Let us construct an algorithm using this shifted polynomial. If the
single update is given by
\begin{align}
r_{n+1} & = r_n - \chi_n A r_n \\
x_{n+1} & = x_n + \chi_n r_n ,
\end{align}
we can generate the solutions $x_n^\sigma$ by
\begin{align} \label{twobeg}
\chi_n^\sigma &= {\chi_n \over 1+\chi_n\sigma} \\
\rho_{n+1}^\sigma &= {\rho_n^\sigma \over 1+\chi_n \sigma}
\label{twoend} \\
x^\sigma_{n+1} &= x^\sigma_n + \chi_n^\sigma \rho_n^\sigma r_n
\end{align}
Remarkably, if $\chi_n$ is generated by the minimal residual condition,
this is exactly the same algorithm which was found in
\cite{m3r} with a completely different approach, namely by a Taylor-expansion
of the residual in $\sigma$ and resummation of the series. This is not
completely surprising, since in the derivation in \cite{m3r} approximations 
were made to achieve that no additional matrix-vector products are needed and
the small recursion length is kept, which automatically leads to the shifted
polynomial. However, the Taylor-expansion becomes prohibitively complex when 
applied to algorithms like BiCGstab, whereas the shifted polynomial method
can easily be transferred. 

\subsection{Three-term recurrences}

Let us now apply these ideas to the case of three-term recurrences, 
which usually appear in algorithms derived from the Lanczos process. 
We look at a general three-term recurrence relation of the form
\begin{equation}
v_{n+1} = \alpha_n A v_n + \beta_n v_n + \gamma_n v_{n-1} \equiv 
Z_{n+1}(A) v_0
\end{equation}
We want to calculate the parameters of the shifted polynomial
\begin{equation}
Z_n^\sigma(z+\sigma) = \zeta^\sigma_n Z_n(A).
\end{equation}
The equations are given by matching the parameters of 
\begin{equation}
\begin{split}
\zeta_{n+1}^\sigma (\alpha_n A v_n + \beta_n v_n + \gamma_n v_{n-1}) = \\
\alpha^\sigma_n A \zeta_n^\sigma v_n + (\beta^\sigma_n+\sigma\alpha^\sigma_n) \zeta_n^\sigma v_n +
\gamma^\sigma_n \zeta_{n-1}^\sigma v_{n-1} .
\end{split}
\end{equation}
The parameters are not completely fixed. One possible choice is
\begin{equation} \label{simple}
\zeta_n^\sigma = 1,\quad \alpha^\sigma_n = \alpha_n, \quad 
\beta^\sigma_n = \beta_n - \sigma\alpha_n, \quad \gamma^\sigma_n = \gamma_n .
\end{equation}
This was realized in \cite{qmr} to construct the QMR and TFQMR method 
for shifted matrices. The Lanczos vectors $v_n$ are in fact independent of 
$\sigma$. If we want to use $v_n$ directly as a residual, we impose the 
condition $\beta_n + \gamma_n = 1$. This determines the parameters of the 
shifted polynomial:
\begin{align}
\alpha^\sigma_n &= \alpha_n {\zeta^\sigma_{n+1} \over \zeta^\sigma_n} \\
\beta^\sigma_n &= (\beta_n - \sigma \alpha_n){\zeta^\sigma_{n+1} \over \zeta^\sigma_n }
\\
\zeta^\sigma_{n+1} &= {\zeta^\sigma_n \zeta^\sigma_{n-1} \over 
\zeta^\sigma_n (1-\beta_n) + \zeta_{n-1}^\sigma (\beta_n-\sigma\alpha_n)} 
\end{align}
with $\gamma^\sigma_n + \beta^\sigma_n = 1$ and 
the initial conditions $\zeta^\sigma_{-1} = \zeta^\sigma_0 = \beta_0 = 1$.
For the case of the Lanczos process it is easy to proof by induction that 
the parameters $\alpha^\sigma_n$,
$\beta^\sigma_n$ and $\gamma^\sigma_n$ are indeed the parameters generated
by the Lanczos process for the matrix $A+\sigma$ if the process does not
break down.
The update of the solution vector is given by
\begin{equation}
x_{n+1}^\sigma = \alpha^\sigma_n \zeta_n^\sigma v_n + \beta^\sigma_n x_n^\sigma + 
\gamma^\sigma_n x_{n-1}^\sigma .
\end{equation}
This is basically the BIORESU$\gamma_5$-algorithm from \cite{borici}. There
the equations (\ref{simple}) are used and an overall normalization factor
is recursively determined. 
It should be noted that this method does not only apply to the Lanczos process,
but for general parameters $\alpha$, $\beta$ and $\gamma$. The shifted 
polynomial will then not be the polynomial generated for the shifted process,
but the shifted systems still converge if $\zeta_i \le 1$.

\subsection{Coupled two-term recurrences}

Now let us turn to the more interesting case of coupled two-term recurrence
relations. These relations have generally a superior numerical stability
compared to the equivalent three-term recurrence. We look at recurrences of 
the CG-type form 
\begin{align}
p_n &= v_n + \alpha_n p_{n-1} = Q_n(A) v_0 \\
v_{n+1} &= v_n + \beta_n A p_n = Z_{n+1}(A) v_0
\end{align}
where the initial condition $p_0 = v_0$ has been used.The method can simply be
applied to a more general choice of parameters. We want to calculate the
parameters needed to generate the 
shifted polynomial $P_n^\sigma(z)$. Unfortunately $Q_n^\sigma$ will 
generally not be a shifted polynomial. This is however not a problem, since
since we can calculate $p_n^\sigma$ without additional matrix-vector products
from
\begin{equation}
p_n^\sigma = \zeta_n^\sigma v_n + \alpha^\sigma_n p^\sigma_{n-1}
\end{equation}
If the vectors $(A+\sigma) p_n^\sigma$ are needed, we can reformulate the
recursion as follows:
\begin{align}
p_n &= v_n + \alpha_n p_{n-1} \\
q_n &= A v_n + \alpha_n q_{n-1} \\
v_{n+1} &= v_n + \beta_n q_n .
\end{align}
We have $q_n = A p_n$ in exact arithmetic. Depending on the algorithm
one or both vectors $p$ and $q$ have to be stored for all values of
$\sigma$. Let us calculate the parameters of the shifted process. To do this,
we derive the three-term recurrence for $v$:
\begin{equation}
v_{n+1} = {\hat\alpha}_n v_n + {\hat\beta}_n A v_n + {\hat\gamma}_n v_{n-1} .
\end{equation}
The parameters are given by
\begin{equation}
{\hat\alpha}_n = \beta_n,\quad {\hat\beta}_n = 
1 + {\beta_n \alpha_n\over \beta_{n-1}}, \quad 
{\hat\gamma}_n = 1 - {\hat\alpha}_n
\end{equation}
with the initial conditions $\alpha_0 = 0$ and $\beta_{-1} = 1$. 
We thus find for the shifted parameters
\begin{align} \label{eqparbegin}
\beta^\sigma_n &= \beta_n {\zeta_{n+1}^\sigma \over \zeta_n^\sigma} \\
\alpha_n^\sigma &= \alpha_n {\zeta_n^\sigma \beta_{n-1}^\sigma \over
\zeta_{n-1}^\sigma \beta_{n-1}} \\
\zeta_{n+1}^\sigma &= {\zeta^\sigma_n \zeta^\sigma_{n-1} \beta_{n-1} \over 
\beta_n \alpha_n(\zeta_{n-1}^\sigma - \zeta_n^\sigma) + \zeta^\sigma_{n-1}\beta_{n-1}
(1-\sigma\beta_n)} 
\label{eqparend}
\end{align}
At the expense of calculating $(A+\sigma)p^\sigma_n$ by introducing
an additional vector and additional dot products, we can also calculate the 
shifted parameters $\beta^\sigma_n$ and $\alpha^\sigma_n$ using the original 
formulae. These formulae do not only apply to the CG process, which will be 
demonstrated below.
We have thus shown that one can implement coupled two-term recurrences for
shifted matrices. 

We can now derive shifted versions of solvers based on these recursion 
relations by simply calculating the shifted parameters and using the
proportionality between the shifted and original polynomials. Whether we
succeed in deriving the shifted algorithm without any additional matrix-vector
products depends on whether matrix-vector products of vectors which are
derived from polynomials which have no shifted structure are needed. In some
cases we can eliminate these matrix-vector products by expressions 
involving other vectors. 

\section{Shifted Krylov space solvers}

In this section we develop shifted algorithm variants of the following
algorithms: CG, CR, BiCG, BiCGstab. In addition shifted versions of the 
solvers QMR, TFQMR and MR are known, so that for most popular Krylov space
methods shifted solvers are available. Note that since TFQMR is based
on CGS, the shifted version of the latter algorithm is basically also 
available. In Table \ref{tab1} we 
present the currently known short recursion methods for shifted matrices
with memory requirements. To avoid a proliferation of new names we propose
to simply add -M to the name of an algorithm to indicate its shifted version.

\begin{table}
\begin{center}
\begin{tabular}{|c|c|c|c|c|}
\hline
Method & Reference & Memory \\
\hline
MR-M & \cite{m3r}, this & $N$ \\
CR-M & this & $2N$ \\
QMR-M (3-term) & \cite{qmr,qmrfromm} & $3N$ \\
QMR-M (2-term) & \cite{qmr2} & $3N$ \\
TFQMR-M & \cite{qmr} & $5N$ \\
BIORESU & \cite{borici}, this & $2N$ \\
BiCG-M & this & $2N$ \\
BiCGstab-M & this & $2N+1$ \\
\hline
\end{tabular}
\end{center}
\caption{\label{tab1} Memory requirements and references for shifted
system algorithms for unsymmetric or nonhermitean matrices. We list the
number of additional vectors neccessary for $N$ additional values of $\sigma$ (
which is independent of the use of the $\gamma_5$-symmetry). }
\end{table}

Note that we cannot easily generalize this method to the CGNE algorithm,
since $(A+\sigma)(A+\sigma)^\dagger$ is not generally a shifted 
matrix. For staggered fermions, however, we are in the lucky position that 
the matrix has the structure
\begin{equation}
D = A + m, \quad A^\dagger = -A
\end{equation}
with $m$ real, so that
\begin{equation}
D D^\dagger = A A^\dagger + m^2
\end{equation}
is a shifted matrix. Since the CG and CR algorithms are optimal for
staggered fermions \cite{deforcrandg5,borici}, we have optimal shifted 
algorithms available for this case.
For Wilson fermions the interesting algorithms are MR and BiCGstab, the 
former due to its simple implementation and small memory requirements
and the latter due to its superior performance and stability; see e.g. 
\cite{frommerlat}.

\subsection{CG-M, BiCG-M, BiCG$\gamma_5$-M}

We present here a version of the CG algorithm for shifted matrices. The variants
BiCG and BiCG$\gamma_5$ are derived analoguosly. Note that the initial guess
has to be set to zero.

\begin{align*}
&{\rm CG-M~~algorithm:} \\
&x_0^\sigma = 0, r_0 = p_0^\sigma = b,
\beta_{-1} = \zeta_{-1}^\sigma = \zeta_0^\sigma = 1, \alpha_0^\sigma = 0 \\
& {\rm for}~~i=0,1,2,\cdots \\
& \quad \beta_i = - {(r_i, r_i)\over (p_i, A p_i)} \\
& \quad {\rm calculate~~} \beta_i^\sigma, \zeta_{i+1}^\sigma 
{\rm~~according~~to~~(\ref{eqparbegin})-(\ref{eqparend})}\\
& \quad x_{i+1}^\sigma = x_i^\sigma - \beta_i^\sigma p_i^\sigma \\
& \quad r_{i+1} = r_i + \beta_i Ap_i \\
& \quad \alpha_{i+1} = {(r_{i+1}, r_{i+1})\over (r_i, r_i)} \\
& \quad {\rm calculate~~} \alpha_{i+1}^\sigma \\
& \quad p_{i+1}^\sigma = \zeta_i^\sigma r_{i+1} + \alpha_i^\sigma p_i^\sigma
\end{align*}
\vskip 0.3cm

This algorithm is a straightforward realization of the formulae 
(\ref{eqparbegin}) - (\ref{eqparend}). Note that we need only 2 additional
vectors for each value of $\sigma$ even in the nonsymmetric BiCG case, since
we can calculate the parameters from the parameters of a single system.
$A$ has to be chosen in a way that $\zeta_i^\sigma \le 1$ for some $i$,
which means that $\sigma = 0$ corresponds usually to the system with the
slowest convergence.

\subsection{CR-M}

The CR algorithm is the truncated version of the generalized conjugate residual
method which is a coupled two-term version of the GMRES
algorithm (see \cite{borici} and references therein). 
We formulate an algorithm which applies the shifted polynomials to the
shifted matrices. 
The algorithm applied to the shifted matrix does in this case not necessarily 
generate the shifted polynomial. The structure is identical to the CG-M
but the parameters are calculated differently, namely we have
\begin{align}
\beta_i &= - {(r_i, A p_i)\over (A p_i, A p_i)} \\
\alpha_{i+1} &= {(A r_{i+1}, A p_i)\over (A p_i, A p_i)} 
\end{align}
Note that formulae (\ref{eqparbegin})--(\ref{eqparend}) still apply although
we do not generate the Lanczos polynomial. 
Note also that we do not know a priori whether this algorithm converges for 
the shifted systems. This has to be checked by testing
\begin{equation}
\zeta_i^\sigma \le 1 .
\end{equation}
If $A$ has only eigenvalues with positive real part, we can however expect that
$\beta$ is generally negative and $\alpha$ positive. If we have
$\zeta_{n-1} > \zeta_n$ we can easily see from formula (\ref{eqparend}) that
$\zeta_n > \zeta_{n+1}$ follows. This suggests that we can expect convergence
if the zero shift corresponds to the system with the worst condition,
which was confirmed in tests with the Wilson fermion matrix.

\subsection{BiCGstab-M}

In the BiCGstab algorithm \cite{bicgstab}, we generate the following sequences
\begin{align}
r_n &= Z_n(A) R_n(A) r_0 \\
w_n &= Z_n(A) R_{n-1}(A) r_0 \\
s_n &= Q_n(A) R_n(A) r_0
\end{align}
where $Z_n(z)$ and $Q_n(z)$ are the BiCG-polynomials and 
\begin{equation}
R_n(z) = \prod_{i=1}^n (1 - \chi_i z) ,
\end{equation}
where the parameters $\chi_i$ are derived from a minimal residual condition.
For the shifted algorithm we have
\begin{align}
r^\sigma_n &= \zeta^\sigma_n \rho^\sigma_n Z_n(A) R_n(A) r_0 \\
w^\sigma_n &= \zeta^\sigma_n \rho^\sigma_{n-1} Z_n(A) R_{n-1}(A) r_0 .
\end{align}
The update of the solution has the form
\begin{equation}
x_{n+1} = x_n - \beta_n s_n + \chi_n w_{n+1} .
\end{equation}
The problem is that the update of $s_n$ itself requires the calculation of
$As_n$, which straightforwardly means we have one additional matrix-vector
multiplication for each value of $\sigma$. But we can use
the relation
\begin{equation}
(A+\sigma)s^\sigma_n = {1\over \beta^\sigma_n} (\zeta^\sigma_{n+1} \rho^\sigma_n w_{n+1} - \zeta^\sigma_n \rho^\sigma_n r_n)
\end{equation}
to eliminate this matrix-vector product at the expense of one auxiliary vector
to store $r_n$. This method is safe since $\beta_n = 0$ only if the
algorithm breaks down anyway. The complete algorithm 
is then given by (note that $s_i \equiv s_i^{\sigma=0}$)

\begin{align*}
& {\rm BiCGstab-M~~algorithm:} \\
&x_0^\sigma = 0, r_0 = s_0^\sigma = b,
\beta_{-1} = \zeta_{-1}^\sigma =  \zeta_0^\sigma =  \rho_0^\sigma = 1,
\alpha_0^\sigma = 0,  \\
& w_0 {\rm ~~so~~that~~} \delta_0 = w_0^\dagger r_0 \ne 0, \phi_0 =
w_0^\dagger A s_0/\delta_0 \ne 0 \\
& {\rm for}~~i=0,1,2,\cdots \\
& \quad \beta_i = - {1\over \phi_i} \\
& \quad {\rm calculate~~}\beta_i^\sigma, \zeta_{i+1}^\sigma
{\rm~~according~~to~~(\ref{eqparbegin})-(\ref{eqparend})}\\
& \quad w_{i+1} = r_i + \beta_i A s_i \\
& \quad \chi_i = { (A w_{i+1})^\dagger w_{i+1}\over (A w_{i+1})^\dagger Aw_{i+1}} \\
& \quad {\rm calculate ~~} \chi_i^\sigma, \rho_{i+1}^\sigma
{\rm~~according~~to~~(\ref{twobeg})-(\ref{twoend})}\\ 
& \quad r_{i+1} = w_{i+1} - \chi_i A w_{i+1}  \\
& \quad x^\sigma_{i+1} = x_i - \beta_i^\sigma s_i + \chi_i^\sigma 
\rho^\sigma_{i}\zeta^\sigma_{i+1} w_{i+1} \\
& \quad \delta_{i+1} = w_0^\dagger r_{i+1} \\
& \quad \alpha_{i+1} = - {\beta_i \delta_{i+1}\over \delta_i \chi_i} \\
& \quad {\rm calculate~~}\alpha_{i+1}^\sigma \\
& \quad s_{i+1} = r_{i+1} + \alpha_{i+1} (s_i - \chi_i A s_i) \\
& \quad s^\sigma_{i+1} = \zeta^\sigma_{i+1} \rho^\sigma_{i+1} r_{i+1} +
        \alpha_{i+1}^\sigma \left(s_i^\sigma - {\chi_i^\sigma \over\beta_i^\sigma } 
(\zeta_{i+1}^\sigma \rho_i^\sigma w_{i+1} - \zeta_i^\sigma \rho_i^\sigma r_i)
\right)
\quad \sigma \ne 0
 \\
& \quad \phi_{i+1} = {w_0^\dagger A s_{i+1} \over \delta_{i+1}}
\end{align*}
\vskip 0.3cm

The convergence of the shifted algorithms 
can be verified by checking that
\begin{equation}
\pi_n \rho_n \le 1 .
\end{equation}
It is however generally advisable for all shifted algorithms to test all 
systems for convergence after
the algorithm finishes since a loss of the condition (\ref{shiftc}) due
to roundoff errors might lead to erratic convergence. 

\section{Preconditioning}

There are two major limitations to shifted algorithms which diminish their 
usefulness considerably. First, we have to start with the same residual for
all values of $\sigma$, which means that cannot have $\sigma$-dependent
left preconditioning. Secondly, preconditioning must retain the shifted
structure of the matrix. While preconditioning can reduce the 
computational effort it has also the important property of numerically
stabilizing the inversion algorithm, which is essential to achieve
convergence in many cases. 

A class of preconditioners which is potentially suitable for shifted systems
are polynomial preconditioners. We note here that we do not expect
to considerably accelerate the matrix inversion algorithms by polynomial
preconditioning in the case of the Wilson matrix since the polynomials
generated by methods like BiCGstab are already nearly optimal
\cite{frommerlat}. We apply a preconditioning polynomial $P_n(z)$ and
solve the equation
\begin{equation}
P_n(A) A y = b, \quad x = P_n(A) y .
\end{equation}
$P_n(z)$ will generally depend on the shift $\sigma$, so we are looking for 
polynomials $P_{n,\sigma}(z)$ which statisfy 
\begin{equation} \label{condi}
P_{n,\sigma}(A+\sigma)(A+\sigma) = P_{n,0}(A)A + \eta 
\end{equation}
and which are good preconditioners. For the linear case, the general solution
is
\begin{equation}\label{linear}
P_{1,\sigma}(z) = 2\sigma+a-z
\end{equation}
where $a$ is an arbitrary constant. 
The case $a=0$ was proposed for the Wilson fermion matrix in \cite{m3r}, 
leading to the preconditioned matrix
\begin{equation}
\left(
\begin{array}{cc}
m^2 - D_{eo}D_{oe} & 0 \\
0 & m^2 - D_{oe} D_{eo} 
\end{array}
\right) ,
\end{equation}
which is fortunately a reasonable preconditioner for the Wilson fermion matrix,
so that the total work is approximately the same as for the unpreconditioned
system. We lose (for general sources) however a factor of two compared to
the usual even-odd preconditioning. 

We assume that generally we do not have to worry too much if
$P_{n,\sigma}(z)$ is a good preconditioner for $\sigma > 0$ 
since usually these systems 
converge faster. Problems only arise if the desired precision is
close to the precision where the residuals stagnate. Given a preconditioner 
of the form 
\begin{equation}
P_n(z) = \prod_{i=1}^n (r_i + z)
\end{equation}
we can calculate the preconditioner 
\begin{equation}
P_{n,\sigma}(z) = \prod_{i=1}^n (r_i^\sigma + z)
\end{equation}
by requiring that (\ref{condi}) holds, which results in a system of $n$ 
equations for the parameters $r_i^\sigma$. Suitable polynomials can 
for example be constructed from Chebychev-, Leja- or GMRES-polynomials.
We will not examine this approach further and only apply linear
preconditioners in our numerical tests.

\begin{figure}
\begin{center}
\epsfig{figure = 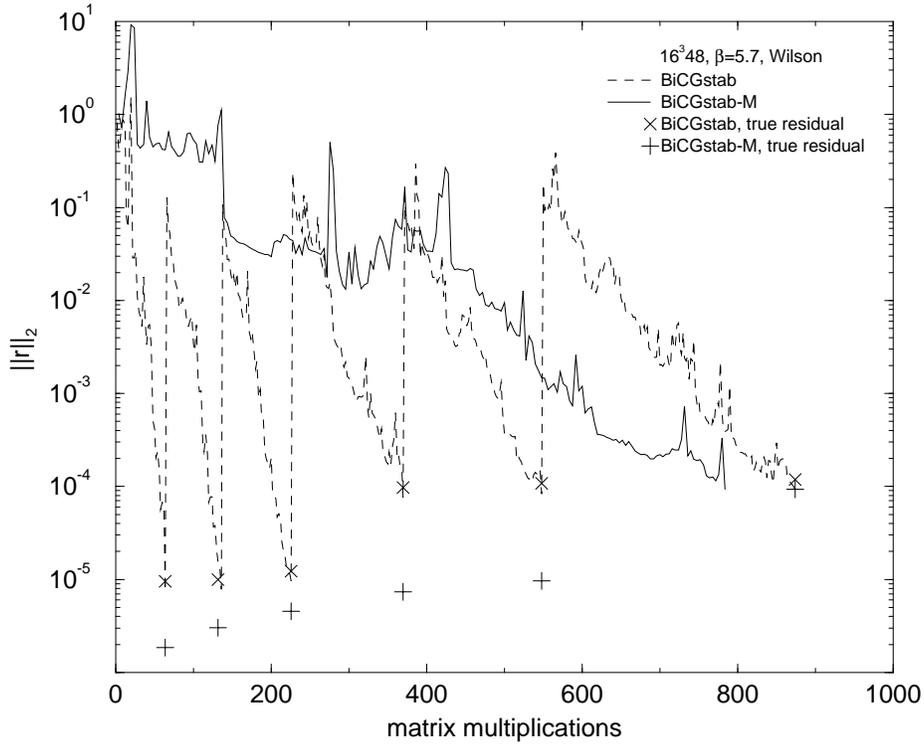, width=120mm}
\end{center}
\caption{\label{fig1} Convergence history for BiCGstab for $\kappa = 0.157$,
0.16, 0.162, 0.165, 0.16625, 0.1675. The BiCGstab algorithm uses even-odd and
BiCGstab-M linear polynomial preconditioning. Note that BiCGstab-M needs
4 matrix multiplications per iteration. }
\end{figure}

\section{Numerical tests}

The algorithms were tested on $16^3\times 48$ quenched $SU(3)$ configurations
at $\beta=5.7$, fixed to Coulomb gauge. 
We used generally 32-bit precision for the vectors and
matrix and 64-bit precision for the accumulation of dot products and 
parameter recursions. The tests were performed on a Cray T3D machine using the
MILC code basis and configurations. 
Other tests of the QMR and MR methods can be found in
\cite{qmrfromm,m3r,compmulti}.

\subsection{Wilson fermions}

The set of hopping parameter values 
was taken from an actual heavy-light calculation with gaussian wall sources. 
We compared the results against t
even-odd preconditioned BiCGstab using the result of lower $\kappa$ values
as initial guesses. We also applied the methods to Clover fermions on the
same configurations.  We performed tests for two lattices,
two spin- and colorindices and sources of size 2 and 6. We found comparable 
results in all cases.

\begin{figure}
\begin{center}
\epsfig{figure = 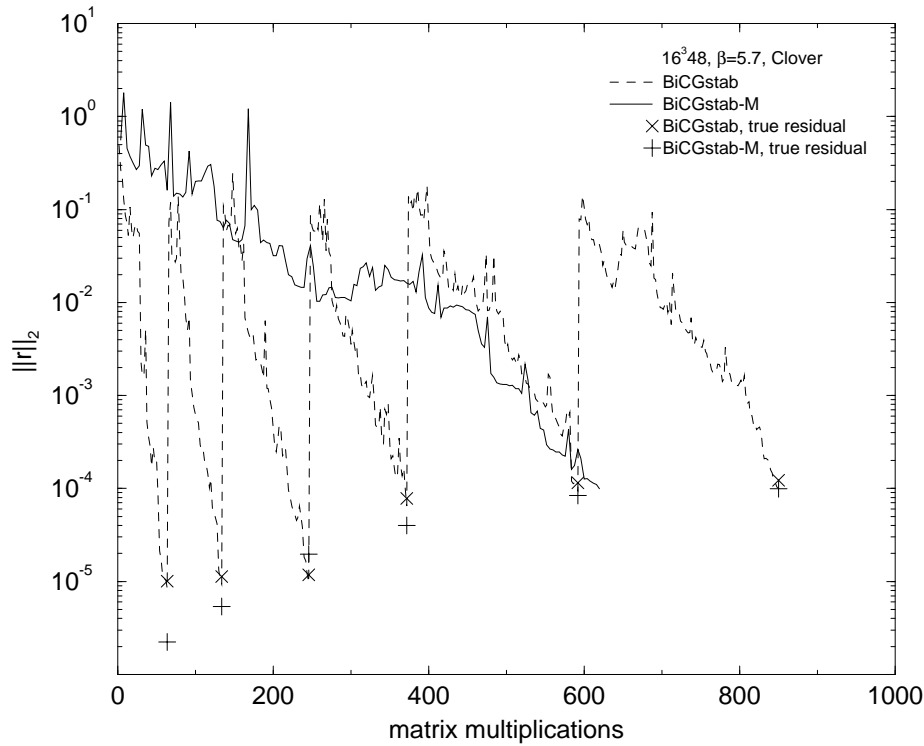, width=120mm}
\end{center}
\caption{\label{figcl1} Convergence history for BiCGstab for $\kappa = 0.136$,
0.38, 0.140, 0.141, 0.142, 0.14235. }
\end{figure}

In Figure \ref{fig1} we show the convergence history of a sample run with 
Wilson fermions taken from an actual production run for heavy-light systems.
The method is (averaged over our test runs) only about 14\% faster than 
BiCGstab with continued guesses, 
which is due to the fact that the gap between the light mass and the
heavier masses is too large. The desired accuracy was $10^{-5}$ for the
3 heavier and $10^{-4}$ for the lighter masses. 
It is easy to see, however, that this factor increases rapidly for mass
values which lie closer together, since the continued guess method cannot
keep the total number of matrix multiplications constant in contrast to
shifted methods. The method is advantageous in a specific case
with $n$ masses (we assume nonlocal sources here), if
\begin{equation}
\sum_{i=1}^n N^{\rm cont}_i - 2 N^{\rm zero~guess}_n \gg 0 .
\end{equation}
The first term is simply the total number of iterations using the standard
algorithm, the last term is twice the number of iterations for the slowest
system using a zero guess. Obviously one can construct examples where this
number can become very large.
Note that for point sources the shifted method wins another
factor 2 in the Wilson fermion case. 
We also tested the MR-M method with an overrelaxation parameter
$\omega = 1.2$. While the MR algorithm performed comparably to the BiCGstab
algorithm in this situation, we found that the residuals of the 
higher mass systems 
stagnate at a value of $\approx 10^{-2}$. This problem was less pronounced
on smaller lattices, so that we assume that it is connected to a loss
of condition (\ref{shiftc}) due to roundoff errors. 
It might however also be connected to our specific implementation.
The same problem can also be seen in \cite{m3r} in Figure 2. 

\subsection{Clover fermions}

We used the tadpole-improved value of the clover constant $c_{SW} = 1.5678$
and values of $\kappa$ so that the inversion takes approximately as long as in
the Wilson case. For the BiCGstab algorithm we used the preconditioned matrix
\begin{equation} \label{ilu}
\left(\begin{array}{cc} 1 & \kappa D_{eo} D_{oo}^{-1} \\
0 & 1 \end{array}\right)
\left(\begin{array}{cc} D_{oo} & -\kappa D_{eo} \\
-\kappa D_{oe} & D_{oo} \end{array}\right)
\left(\begin{array}{cc} 1 & 0 \\
\kappa D_{oo}^{-1} D_{eo} & D_{oo}^{-1} \end{array}\right) .
\end{equation}
For BiCGstab-M we used the linear preconditioner (\ref{linear}) with $a=0$.
The preconditioned matrix does not separate nicely like in the Wilson case,
which makes however no difference in the computational effort
for general sources.
It does however serve its main purpose, namely to stabilize the algorithm
sufficiently so that it converges in our test cases. We find that the
implementation of the preconditioner is important in the sense that
a violation of condition (\ref{condi}) due to roundoff errors can lead
to a stagnation of the shifted residuals. 
The number of iterations
needed with zero initial guess 
is approximately the same for the BiCGstab and BiCGstab-M for the
smallest mass which means that the linear preconditioner reduces the condition
of the matrix as well as the preconditioner (\ref{ilu}).
The further conclusions are therefore similar to the Wilson fermion 
case. In Figure \ref{figcl1} we show a convergence history for a system with
clover fermions. 
Note that we saw examples of a loss of precision in the shifted 
residuals which lead to early stagnation, so that it is advisable to check the
residuals of the shifted systems for convergence. Here the mass values lie
effectively closer together and a bigger improvement can be seen.

\section{Conclusions}

We presented a simple point of view to understand the structure of
Krylov space algorithms for shifted systems, allowing us to construct
shifted versions of most short recurrence Krylov space algorithms.
We developed the shifted CG-M and CR-M algorithm which can be applied
to staggered fermion calculations. Since efficient 
preconditioners for the staggered fermion matrix are not known, 
a very large improvement by these algorithms can be expected.
We also presented the BiCGstab-M 
method, which, among the shifted algorithms, 
is the method of choice for quark propagator calculations
using Wilson (and presumably also Clover) fermions if enough memory 
is available. It becomes available simply by extending existing BiCGstab 
implementations. 
We investigated the
efficiency of this method in realistic applications and found that,
for sources other than point sources, the improvement depends heavily on the
values of the quark masses. The improvement is generally higher for masses
which lie closer together. 
The numerical stability of convergence of the
shifted systems is found to be very good so that this method is feasible
in 32-bit arithmetic. The application of this method to Clover fermions is
possible. Using simple linear polynomial preconditioning we can stabilize the
solver sufficiently even for relatively small quark masses. We proposed a 
way to apply higher order polynomial preconditioners to shifted matrix solvers
which may be neccessary in the case of very small quark masses. 
Roundoff errors might however in some cases affect the convergence of the
shifted systems so that the final residuals have to be checked. 

\section*{Acknowledgements}

This work was supported in part by the U.S. Department
of Energy under Grant No. DE-FG02-91ER40661. Computations were performed on
the Cray T3D at the San Diego Supercomputer Center.
I would like to thank S. Pickels, C. McNeile and S. Gottlieb for
helpful discussions.


\begin{thebibliography}{99}

\bibitem{qmr} R. W. Freund, Solution of shifted linear systems by quasi-minimal
residual iterations, Numerical Linear Algebra, L. Reichel, A. Ruttan and
R.S. Varga (eds.), Berlin: W. de Gruyter 1993, 101--121.

\bibitem{deforcrandg5} P. de Forcrand, Progress on lattice QCD algorithms, IPS-95-24, hep-lat/9509082

\bibitem{m3r} U. Gl\"assner, S. G\"usken, T. Lippert, G. Ritzenh\"ofer,
K. Schilling and A. Frommer, 
How to compute Green's functions for entire mass trajectories within
Krylov solvers, hep-lat/9605008

\bibitem{borici} A. Bori\c{c}i, Krylov subspace methods in lattice QCD, SCSC report
TR-96-27

\bibitem{leja} L. Reichel, The application of Leja points to Richardson
iteration and polynomial preconditioning, Linear Algebra and its Applications 145--156 (1991)
389

\bibitem{qmr3} R.W. Freund, M. H. Gutknecht, N.M. Nachtigal, An implementation
of the look-ahead lanczos algorithm for non-Hermitean matrices, 
SIAM J. Sci. Comput. 14 (1993) 137

\bibitem{qmr2} R.W. Freund and N.M. Nachtigal, An implementation of the {QMR} method based on coupled two-term recurrences, http://cm.bell-labs.com/cm/cs/doc/92/4-06.ps.gz


\bibitem{frommerlat} A. Frommer, Linear system solvers: Recent developments and 
implications for lattice computations, to appear in the proceedings of 
the Lattice '96 conference, hep-lat/9608074

\bibitem{bicgstab} H.A. van der Vorst, Bi-CGstab: A fast and smoothly converging Variant 
of Bi-CG for the solution of nonsymmetric linear systems, SIAM J. Sci Statist. Comput. 13 (1992) 631,
M. H. Gutknecht, Variants of BiCGstab for matrices with complex spectrum,
SIAM J. Sci. Comput. 14 (1993) 1020

\bibitem{qmrfromm} A. Frommer, B. N\"ockel, S. G\"usken, T. Lippert and K. Schilling, 
Many masses on one stroke: Economic computation of quark propagators, 
HLRZ-95-20, hep-lat/9504020

\bibitem{compmulti} H-P. Ying, S-J. Dong and K-F. Liu, Comparison of
multi-quark
matrix inversion algorithms, hep-lat/9611009


\bibitem{gupta} R. Gupta, T. Bhattacharya and G. Kilcup, Comparison of inversion algorithms for Wilson fermions on the CM5, LA-UT-96-1115, hep-lat/9605029


\end{thebibliography}
\end{document}